\documentclass[useAMS,usenatbib]{mn2e}
\usepackage{graphicx,amssymb,amsmath}
\usepackage{mnras_macros}
\usepackage[dvips]{color}

\newcommand{\bm}[1]{\boldsymbol  #1 }
\newcommand{\dm}{{\rm d}}
\newcommand{\ion}[2]{#1$\,${\sc #2}}
\newcommand{\HI}{\ensuremath{\mbox{\ion{H}{i}}}} 
\newcommand{\HII}{\mbox{\ion{H}{ii}}} 
\newcommand{\HeI}{\mbox{\ion{He}{i}}} 
\newcommand{\HeII}{\mbox{\ion{He}{ii}}} 
\newcommand{\HeIII}{\mbox{\ion{He}{iii}}} 

\title[ARGOT: Accelerated radiative transfer]
{ARGOT: Accelerated radiative transfer on grids using oct-tree}
\author[T. Okamoto, K. Yoshikawa, \& M. Umemura]{Takashi
Okamoto$^{1}$\thanks{E-mail: 
tokamoto@ccs.tsukuba.ac.jp}, Kohji Yoshikawa$^{1}$, 
Masayuki Umemura$^{1}$\\ 
$^{1}$ Center for Computational Sciences, University of Tsukuba, 1-1-1 
 Tennodai, Tsukuba 305-8577 Ibaraki, Japan  
}
\begin{document}

\date{Accepted . Received ; in original form }

\pagerange{\pageref{firstpage}--\pageref{lastpage}} \pubyear{2009}

\maketitle

\label{firstpage}

\begin{abstract}
We present two types of numerical prescriptions that accelerate the radiative 
transfer calculation around point sources within a three-dimensional 
Cartesian grid by using the oct-tree structure for the distribution of 
radiation sources. 
In one prescription, distant radiation sources are grouped as 
a bright extended source when the group's angular size, $\theta_{\rm s}$, 
is smaller than a critical value, $\theta_{\rm crit}$, and radiative 
transfer is solved on supermeshes whose angular size is similar to 
that of the group of sources. 
The supermesh structure is constructed by coarse-graining the mesh 
structure. 
With this method, the computational time scales with 
$N_{\rm m} \log(N_{\rm m}) \log(N_{\rm s})$ where $N_{\rm m}$ and 
$N_{\rm s}$ are the number of  meshes and that of radiation sources, respectively. 
While this method is very efficient, it inevitably overestimates 
the optical depth when a group of sources acts as an extended powerful 
radiation source and affects distant meshes. 
In the other prescription, a distant group of sources is treated as 
a bright point source ignoring the spatial extent of the group 
and the radiative transfer is solved on the meshes rather than the 
supermeshes. 
This prescription is simply a grid-based version of {\scriptsize START} 
by Hasegawa \& Umemura and yields better results in general with 
slightly more computational cost ($\propto N_{\rm m}^{4/3} \log(N_{\rm s})$) 
than the supermesh prescription. 
Our methods can easily be implemented to any grid-based hydrodynamic codes 
and are well-suited to adaptive mesh refinement methods. 
\end{abstract}


\section{Introduction}

Radiative transfer (RT) of photons has fundamental importance for 
formation of astronomical objects, such as galaxies, stars, and 
blackholes. Unfortunately, the nature of RT, 
{in which we have to solve the time evolution of the six-dimensional 
 phase-space information of photons} 
(three spatial dimensions, two angular dimensions, and one 
frequency dimension; or equivalently three spatial and three 
momentum dimensions), makes it difficult to solve RT accurately and 
to couple it with hydrodynamics. To date, various RT schemes has been 
proposed \citep{iliev06}, some of which are coupled with hydrodynamics
\citep{iliev09}.  
A wide range of approximation have been used to deal with multi-dimensional 
nature of the transfer equation and they have their own pros and cons. 

When radiation sources are embedded in media on meshes,
RT calculations can be categorised into two types;
one premises that the source functions are assigned on meshes
and the other does that radiation sources are treated as
point sources independent of meshes.
In the former type, the RT equations are integrated
along long or short characteristics between meshes.
The latter is advantageous when the number of the point sources, 
$N_{\rm s}$, is smaller than that of the boundary meshes, 
$\sim N_{\rm m}^{2/3}$, where $N_{\rm m}$ is the total number of the 
meshes.  
The latter type of the RT schemes is often called `ray-tracing' that we 
deal with in this paper.  

The most accurate and straight-forward RT scheme is the long characteristics
method in which all source meshes are connected to all other relevant meshes 
\citep{abel99, sokasian01, RSPH}.  
This method is however very expensive computationally. 
The computational costs scales with $N_{\rm m}^2$ in general and 
with $N_{\rm m}^{4/3} N_{\rm s}$ for the transfer from point sources. 

The short characteristics method \citep{kunasz88, stone92, mellema98, nakamoto01} 
reduces the computational cost by integrating the equation of RT only along lines that 
connect nearby cells. 
It scales with $N_{\rm m}^{5/3}$ and with $N_{\rm m} N_{\rm s}$ for the transfer 
from point sources. Its known disadvantage is the inability to track collimated 
radiation fields and hence the inability to cast sharp shadows owing to the 
numerical diffusion. 

The methods whose computational cost is similar to that of the short characteristics method 
with small loss of accuracy compared to the long characteristics method have also 
been developed (\citealt{razoumov05} and `authentic RT' by Nakamoto et al. in 
\citealt{iliev06}). 
Adaptive ray tracing \citep{abel02} has been widely used for RT around point sources
\citep{wise11}.  

Mote Carlo transport \citep{ciardi01} is also straight forward. The advantage of this
approach is that comparatively few approximations to the RT equations 
need to be made. The resulting radiation field however inevitably becomes noisy 
\citep[see][]{iliev06} due to its stochastic nature unless a huge number of photon packets 
are transported. This method is computationally very expensive in the optically 
thick regime.

The methods,  which consider the moments of the RT equations and consist in choosing a closure 
relation to solve them, can lead to substantial simplifications that can drastically 
speed up the calculations because its computational cost scales with $\sim N_{\rm m}$. 
The most common of these methods is the flux-limited diffusion, which solves the evolution 
of the first moment and uses a closure relation valid in the diffusion limit, 
which is an isotropic radiative pressure tensor. The equation is modified with an ad-hoc
function (the flux limiter) in order to ensure that the radiative flux is valid in 
the free-streaming limit. This method is very useful in diffusive regions and have been used to 
study accretion discs \citep{ohsuga05} and star formation \citep{krumholz06}. 
Another method of closing the system is the variable Eddington tensor formalism. 
It gives better results than the flux-limited diffusion but are much more complex and 
costly because it requires the local resolution of the transfer equation at each timestep. 
The methods which employ the optically thin variable Eddington tensor approximation 
\citep{ga01} have been used to study cosmic reionization \citep{ga01, ricotti02, petkova09}. 
A locally evaluated Eddington tensor, called the M$_1$ model, has also been used to close the 
system \citep{heracles} and has applied to study  cosmic reionization \citep{aubert08}. 
The accuracy of the moments methods is problem-dependent and is hard to judge in general 
situation.  
\citet{petkova11} have developed a method that employs a direct discretisation of 
the RT equation in Boltzmann form with finite angular resolution on moving meshes. 
This method is advantageous in solving problems in which time-dependent solution of the RT 
equation is important. The timestep however has to be very short because 
photons propagate at the speed of light unless a reduced speed of light approximation 
is employed. 

In many astrophysical problems, for example cosmic reionization and galaxy formation, 
we have to deal with numerous radiation sources. 
{
\citet{TRAPHIC} introduced source merging procedure in order to avoid computationally 
  expensive scaling with the number of sources and implemented it on Smoothed Particle 
  Hydrodynamics (SPH). 
}
\citet{START} utilised the oct-tree algorithm \citep{tree} in order to 
accelerate the RT around point sources and they coupled the RT with SPH. 
In their method, distant sources from a target gas particle are grouped 
and regarded as a single point source when the angular size of the group of the 
sources is smaller than a critical value. 
Consequently, the effective number of radiation sources is largely reduced to 
$\log(N_{\rm s})$ when there are $N_{\rm s}$ sources. 

The methods we explore in this paper are parallel to this approach except that 
we implement this grouping algorithm to grid-based codes. 
In one of our methods, we introduce supermeshes; a supermesh consists of $8^n$ 
meshes and it is characterised by the mean density of each chemical species of the 
meshes within the supermesh. Solving the RT on supermeshes whose angular size is 
similar to that of the group of the sources in question results in further reduction 
of computational time in principle. 
Another approach we take is the point source approximation, in which a group of 
sources sufficiently distant from a target mesh is treated as a point source. 
The latter can be regarded as a grid-based version of {\scriptsize START} \citep{START}. 

Unlike gravitational interactions to which the tree-algorithm has been widely  
applied, RT is affected by the medium between a source and a target.  
It is therefore very important to test these tree-based approaches 
in cases where an extended group of sources works as a powerful source 
in inhomogeneous medium and affects (e.g. ionizes) distant meshes.  
In this paper, we extensively investigate such cases in order to clarify 
advantages and disadvantages of the methods using tree-based algorithm. 

This paper is organised as follows. In section 2, we describe the algorithm 
in detail. In section 3, we present several test problems and compare 
our methods to each other.
We summarise and discuss the results in section 4. 

\section{Radiative transfer with tree-algorithm}

In this section, we describe our ray-tracing algorithm that 
we use to solve the steady RT equation for a given frequency, $\nu$: 
\begin{equation}
\frac{\dm I_\nu}{\dm \tau_\nu} = - I_\nu + S_\nu, 
  \label{eq:rt}
\end{equation}
where $I_\nu$, $\tau_\nu$, and $S_\nu$ are the specific intensity, 
the optical depth, and the source function, respectively. 
This equation is adequate for problems in which the absorption 
and emission coefficients change on timescales much longer than 
the light crossing time. This will always be the case in the volumes 
we will simulate by using our methods. 
Eqn.~(\ref{eq:rt}) has a formal solution: 
\begin{equation}
I_\nu(\tau_\nu) = I_{\nu, 0} \mathrm{e}^{- \tau_\nu} 
  + \int_0^{\tau_\nu} S_\nu(\tau'_\nu) \mathrm{e}^{-\tau_\nu + \tau'_\nu} \dm \tau'_\nu, 
\label{eq:formal}
\end{equation}
where $I_{\nu, 0}$ is the specific intensity at $\tau_\nu = 0$ and 
$\tau'_\nu$ is the optical depth at a position along the ray. 
Throughout this paper we employ so-called on-the-spot approximation 
\citep{agnagn} in which recombination photons are assumed to be absorbed 
where they were emitted. 
Using the on-the-spot approximation, the formal solution given by equation 
(\ref{eq:formal}) is reduced to 
\begin{equation}
I_\nu(\tau_\nu) = I_{\nu, 0} \mathrm{e}^{- \tau_\nu}.  
\label{eq:formal2}
\end{equation}
To solve this equation numerically, one needs to calculate optical depth 
between each pair of a source and a target mesh. 
The computational cost is hence proportional to the number of sources. 
In the next subsection, we will describe the method to decrease the effective 
number of radiation sources by using the oct-tree structure. 

\subsection{Source grouping algorithm}

As in \citet{START}, we construct the oct-tree structure for the 
distribution of radiation sources. 
A cubic computational domain is hierarchically subdivided into 8 
cubic cells until each cell contains only one radiation source or the 
size of a cell becomes sufficiently small compared to that of the 
computational domain. 
We call these sub-volumes `tree nodes'. 
When the side length of the cubic computational domain is $L$, the width 
of a level $l$ tree node is given by $w^{(l)} = L/2^{l}$. 
Each tree node records the centre of the luminosity of the radiation 
sources contained in the node,
\begin{equation}
\bm{r} = \frac{\sum_m \bm{r}_m L_m}{\sum_m L_m}, 
\label{eq:centre}
\end{equation}
and the total luminosity, 
\begin{equation}
L = \sum_m L_m,  
\label{eq:totallum}
\end{equation}
where $\bm{r}_m$ and $L_m$ indicate the position vector and the luminosity 
of a radiation source, respectively, and subscript $m$ runs over all sources 
within the tree node. 

Once we have constructed the tree structure, we loop over all meshes. 
RT from all the radiation sources to each target 
mesh is performed by a simple recursive calculation as done in 
$N$-body calculation. 
We start at the root node (level 0 tree node), which covers entire 
computational domain. 
Let $w$ be the width of the node currently being processed and $D$ 
the distance between the closest edges of the tree node and the 
target mesh. If the angular size of the node is smaller than a 
fixed value of accuracy parameter, i.e. 
\begin{equation}
\frac{w}{D} < \theta_{\rm crit}, 
  \label{eq:opening}
\end{equation}
then we perform the RT calculation between the group of sources 
in the current node and the target mesh and move on to the next node. 
Otherwise, we examine the child nodes (subnodes) recursively. 
The effective number of sources is thus proportional to $\log(N_{\rm s})$. 
In the following subsections we will explain how we perform 
the RT calculation between a group of sources and a target mesh. 

\subsection{Supermesh approximation} \label{sec:smart}
\begin{figure}
\begin{center}
\includegraphics[width=8.0cm]{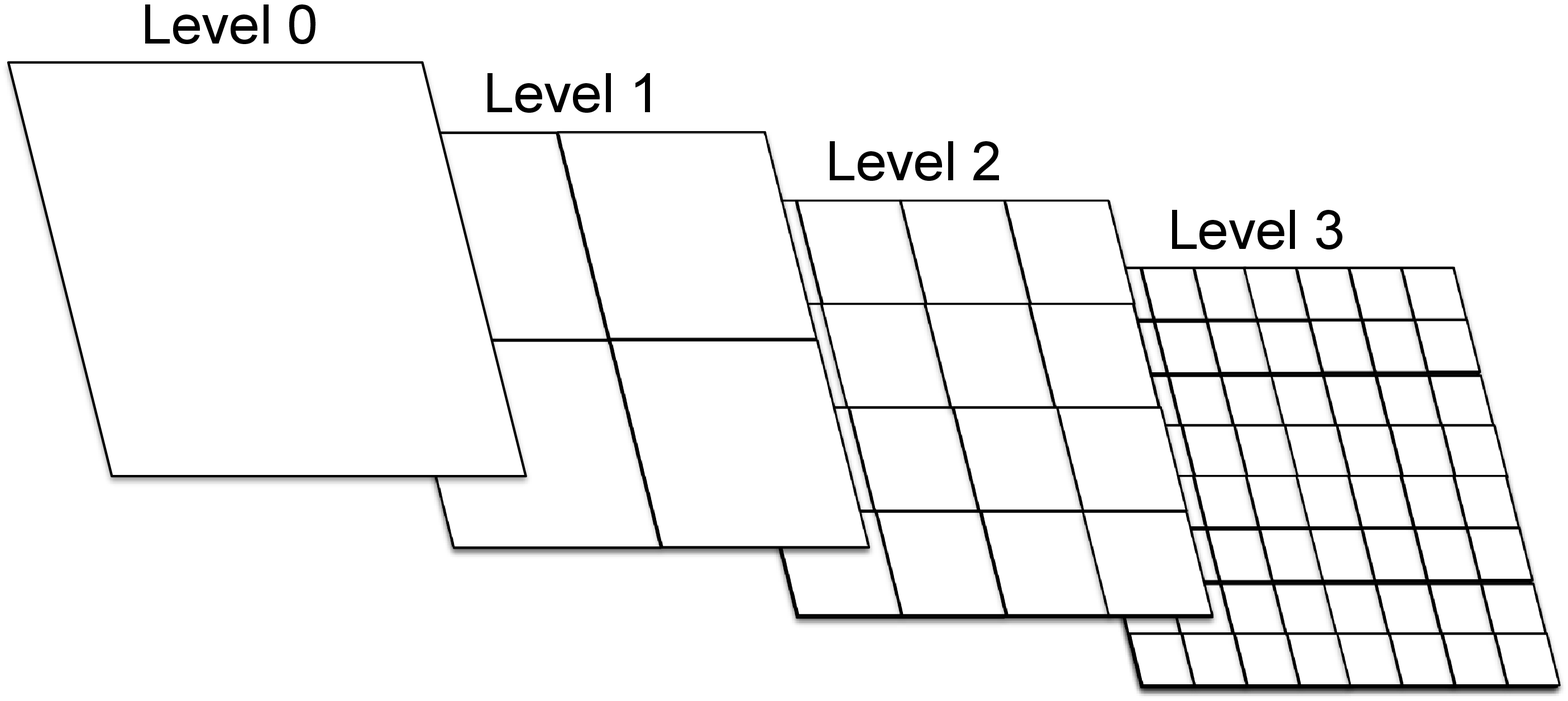}
\end{center}
\caption{Schematic illustration of the supermesh structure for $8\times8$ 
  two-dimensional meshes. In this case, the maximum level, $l_{\rm max}$, is 3 
  and the meshes themselves can be used as the highest level supermeshes. 
  A level $l$ supermesh contains $2^{2(l_{\rm max} - l)}$ meshes. 
  For three-dimensional meshes, a level $l$ supermesh consists of 
  $2^{3(l_{\rm max} - l)}$ meshes. 
}
\label{fig:supermesh}
\end{figure}
We first introduce the supermesh approximation. 
In Fig.~\ref{fig:supermesh}, we show a schematic illustration of the supermesh 
structure. If a three-dimensional computational domain is discretised by 
$2^{3 l_{\rm max}}$ meshes, a level $l$ supermesh consists of  
$2^{3 (l_{\rm max} -l)}$ meshes.  
We can calculate the mean density of each chemical species for 
every supermesh by using the meshes contained in it. 
The meshes can be used as the highest level supermeshes. 
The supermesh structure is resembling to an adaptive mesh refinement (AMR) 
structure and thus this method is well-suited to couple with the hydrodynamics
by AMR codes. 

\begin{figure}
\begin{center}
\includegraphics[width=8.0cm]{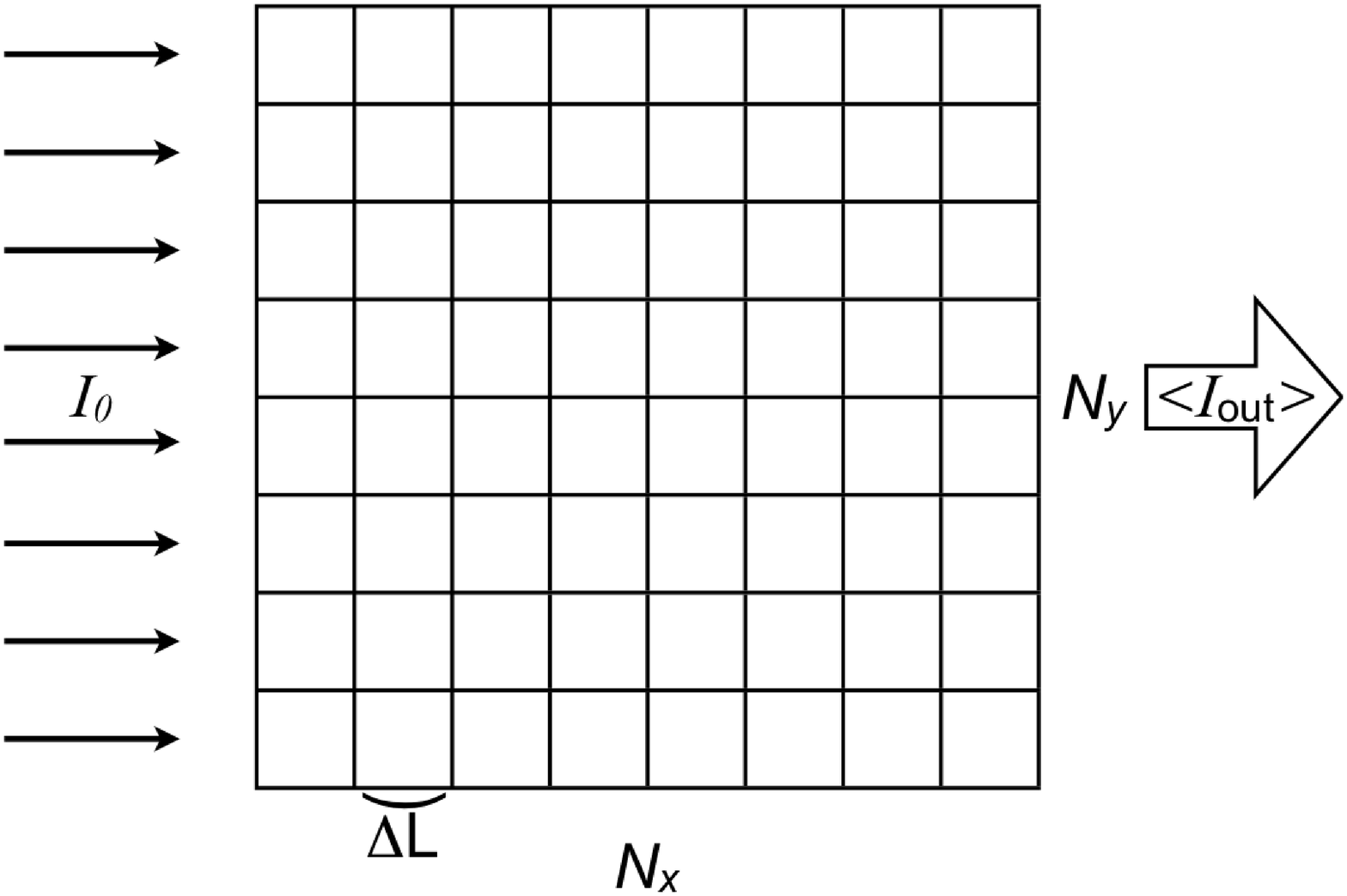}
\end{center}
\caption{Plane-parallel radiation with specific intensity $I_0$ entering 
  to a supermesh that consists of $N_x \times N_y$ meshes. 
  The $(i, j)$-th mesh has the $\HI$ number density, $n_{i, j}$. 
}
\label{fig:supermeshapprox}
\end{figure}

Let us consider the case in which plane-parallel radiation 
with the specific intensity $I_0$ enters a supermesh that 
consists of $N_x \times N_y$ meshes. 
What we want to know is the mean intensity of the ray emerging from 
the other side of the supermesh, $\langle I_{\rm out} \rangle$ 
(see Fig.~\ref{fig:supermeshapprox}). 
For simplicity, we here only consider the absorption by $\HI$ atoms and drop 
the frequency dependence. The side length of each mesh is $\Delta L$ and 
the $\HI$ number density of the $(i, j)$-th mesh in the supermesh 
is $n_{i, j}$. 
The mean intensity of the emerging radiation is given by 
\begin{equation}
\langle I_{\rm out} \rangle = \frac{I_0}{N_y} \sum_j^{N_y} \exp[- \sigma_{\rm HI} {\cal N}_j], 
  \label{eq:exact}
\end{equation}
where $\sigma_{\rm HI}$ is the $\HI$ cross-section and ${\cal N}_j$ is the 
$\HI$ column density of the $j$-th line, i.e. ${\cal N}_j = \sum_i^{N_x} n_{i,j} \Delta L$. 

{
In our supermesh approximation, we use the mean $\HI$ number density, 
   $\langle n \rangle = \sum_{i,j} n_{i, j} / (N_x N_y)$, 
to estimate the mean intensity of the emerging radiation 
$\langle I_{\rm out} \rangle$. 
Doing this introduces some error as we will show below. 
In order to understand the accuracy and the nature of the supermesh 
approximation, we compare the mean intensity of the emerging radiation 
by the supermesh approximation to that calculated by using the meshes. 
We first consider the Taylor series expansion of the mean 
intensity of the emerging radiation when we solve the RT on the supermesh: 
}
\begin{eqnarray}
\langle I_{\rm out} \rangle_{\rm mean} &=& I_0 
  \exp(-\sigma_{\rm HI}\langle {\cal N} \rangle ) \nonumber \\
  & = & I_0 \left[
  1 - \sigma_{\rm HI} \langle {\cal N} \rangle 
  + \frac{\sigma_{\rm HI}^2}{2} \langle {\cal N} \rangle^2  + \cdots \right], 
  \label{eq:mean}
\end{eqnarray}
where $\langle {\cal N} \rangle$ is the mean $\HI$ column density given by 
\begin{eqnarray}
\langle {\cal N} \rangle = \frac{\sum^{N_y}_j {\cal N}_j}{N_y} 
=  \frac{\Delta L \sum_j^{N_y} \sum_i^{N_x} n_{i, j}}{N_y} = \Delta L N_x \langle n \rangle. 
\label{eq:column}
\end{eqnarray}

On the other hand, the Taylor series expansion of Eqn.~(\ref{eq:exact}) is 
\begin{eqnarray}
\langle I_{\rm out} \rangle &=& \frac{I_0}{N_y} 
\sum_j^{N_y} \left[1 - \sigma_{\rm HI} {\cal N}_j + \frac{1}{2} (\sigma_{\rm HI} {\cal N}_j)^2  
+ \cdots \right] \nonumber  \\
    &=& \frac{I_0}{N_y} \left[N_y - \sigma_{\rm HI} \sum^{N_y}_j {\cal N}_j 
    + \frac{\sigma_{\rm HI}^2}{2} \sum^{N_y}_j {\cal N}_j^2  + \cdots \right]  \nonumber \\
    &=& I_0 \left[1 - \sigma_{\rm HI} \langle {\cal N} \rangle 
    + \frac{\sigma_{\rm HI}^2}{2} \langle {\cal N}^2 \rangle + \cdots \right].  
\label{eq:tayler}
\end{eqnarray}
The difference between $\langle I_{\rm out} \rangle$ and $\langle I_{\rm out} \rangle_{\rm mean}$ is 
the second order in $\tau$. From Eqn.~(\ref{eq:exact}) and (\ref{eq:mean}), 
the leading error in $\langle I_{\rm out} \rangle_{\rm mean}$ is
\begin{eqnarray}
\langle I_{\rm out} \rangle - \langle I_{\rm out} \rangle_{\rm mean} &=& I_0 \frac{\sigma_{\rm HI}^2}{2}
  \left(\langle {\cal N}^2 \rangle  - \langle {\cal N} \rangle^2 \right). 
  \label{eq:error}
\end{eqnarray}
Since the variance of the column density, 
$\langle {\cal N}^2 \rangle - \langle {\cal N} \rangle^2$, could be very large in 
the inhomogeneous medium, we substantially overestimate the optical depth 
if we use Eqn.~(\ref{eq:mean}).

We can therefore in principle improve the approximation by estimating 
the variance of the column density. 
According to the central limit theorem, the variance of the column density 
for large $N_x$ can be expressed by using the variance of the density, 
if the density, $n_{i, j}$, is a sequence of independent and identically 
distributed random variables:  
%
\begin{equation}
\langle {\cal N}^2 \rangle - \langle {\cal N} \rangle^2 = N_x \left[\langle n^2 \rangle - \langle n \rangle^2\right]. 
\label{eq:centrallimit}
\end{equation}
Using this relation, the mean intensity of the emerging radiation can be approximated as: 
\begin{eqnarray}
\langle I_{\rm out} \rangle_{\rm variance} &=& I_0 \biggl[\exp\left(-\sigma_{\rm HI} \langle {\cal N} \rangle \right) \nonumber \\
    && \hspace{1cm} + \frac{\sigma_{\rm HI}^2 N_x}{2}\left(\langle n^2 \rangle - \langle n \rangle^2\right)\biggl]. 
    \label{eq:smapprox}
\end{eqnarray}
The effective column density for a ray segment that intersects the supermesh is hence 
\begin{equation}
{\cal N}_{\rm eff} = - \frac{\ln\left[\exp\left(-\sigma_{\rm HI} \langle n \rangle h\right) 
+ \frac{\sigma_{\rm HI}^2 h}{\Delta L} \left(\langle n^2 \rangle - \langle n \rangle^2\right)\right]}{\sigma_{\rm HI}}, 
\label{eq:effectivecolumn}
\end{equation}
where $h$ is the length of a ray segment. 
We however do not employ this approximation 
because Eqn.~(\ref{eq:centrallimit}) is only valid for large $N_x$ and $N_x$ always 
becomes small near the target mesh. 
We thus only use the mean density in our supermesh approximation which is
described by Eqn.~(\ref{eq:mean}).  
We will investigate the accuracy of this approximation in Section~\ref{sec:tests}. 

\begin{figure}
\begin{center}
\includegraphics[width=8.0cm]{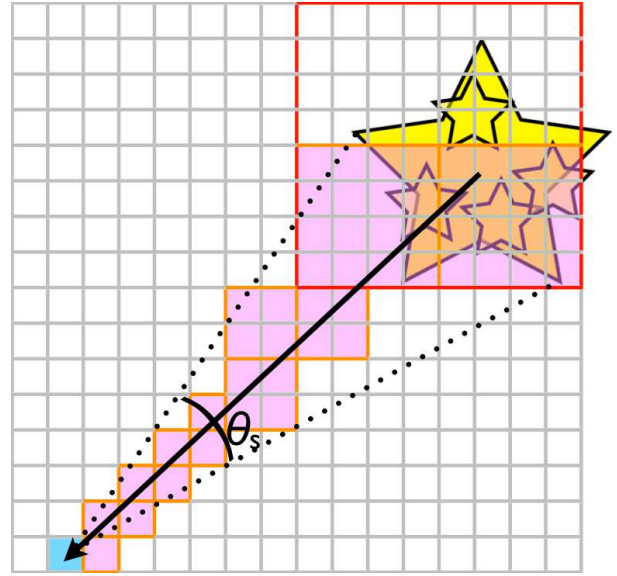}
\end{center}
\caption{A schematic illustration of the RT calculation using supermeshes. 
  The radiation sources in the tree node indicated by red square are regarded 
  as a single bright extended source. 
  The target mesh is coloured by light blue. 
  The RT is solved on the supermeshes at 
  the lowest level, whose angular size is equal to or smaller than the angular 
  size of the source group, $\theta_{\rm s}$. The supermeshes used are indicated 
  by purple colour and their sizes are represented by the orange squares. 
}
\label{fig:supermeshrt}
\end{figure}

Now we have to determine on which supermeshes we perform the RT calculation. 
We chose to use the lowest level supermeshes whose angular size, $\theta$,  
is equal to or smaller than the angular size of a group of the sources, $\theta_{\rm s}$, 
since we assume plane-parallel radiation to construct the approximation. 
We define the luminosity-weighted rms projected radius as the effective 
projected size of the group of the sources\footnote{This choice may somewhat 
underestimate the effective projected size as for the case of a disc with a 
constant surface brightness. We have confirmed that simulation results are not 
sensitive to such a level of difference (a factor of $\sqrt{2}$).}, 
i.e. if the target mesh is located along the $z$-direction from the 
centre of the luminosity, the projected size of the group is defined as 
\begin{equation}
r_{\rm rms}^2 = \frac{\sum_m L_m \left\{(x_m - \bar{x})^2 + (y_m - \bar{y})^2\right\}}{\sum_m L_m}, 
  \label{eq:rms}
\end{equation}
where $\bar{x}$ and $\bar{y}$ are, respectively, the $x$ and $y$ components of 
the position vector of the luminosity centre and the subscript $m$ runs over 
all sources in the tree node in question.  
Practically, we calculate the following tensor for each tree node: 
\begin{equation}
{\cal I}_{ij} = \sum_m L_m (\bm{r}_{m, i} - \bar{\bm{r}}_i)(\bm{r}_{m, j} - \bar{\bm{r}}_j), 
\label{eq:inertia}
\end{equation}
where the subscripts $i$ and $j$, respectively, indicate $i$-th and $j$-th components of the position 
vector, i.e. $i$ and $j$ are either $x$, $y$, or $z$; and the subscript $m$ has the same meaning as in 
Eqn.~(\ref{eq:rms}). 
By using $(0, 0)$ and $(1, 1)$ components of the tensor ${\cal I}'$  which is the 
tensor ${\cal I}$ in the rotated frame so that the target mesh is placed along the $z$-direction 
from the luminosity centre, we can estimate the angular size of the group of the source in 
the tree node as 
\begin{equation}
\theta_{\rm s} = \frac{2 r_{\rm rms}}{D} =  \frac{2}{D} \left(\frac{{\cal I}'_{00} +  {\cal I}'_{11}}{\sum_m L_m} \right)^\frac{1}{2}, 
\label{eq:angularsize}
\end{equation}
where $D$ is the distance between the luminosity centre and the closest edge of the target mesh. 
In Fig.~\ref{fig:supermeshrt}, we illustrate the procedure of the RT calculation using the 
supermeshes. 
The computational cost by this method is expected to scale 
with $N_{\rm m}  \log(N_{\rm m}) \log(N_{\rm s}) $. 

\subsection{Point source approximation} \label{sec:start}

Here we introduce another way of accelerating the RT calculation by 
using the oct-tree structure of the distribution of radiation sources. 
As in \citet{START}, we treat a group of sources in a tree node which satisfies 
the condition described by Eqn.~(\ref{eq:opening}) as a bright point source. 
Since we ignore the size of the source group, we solve the RT not on the supermeshes 
but on the meshes. Consequently, the computational cost scales with 
$N_{\rm m}^\frac{4}{3} \log(N_{\rm s})$.
Although this is slightly more expensive computationally than the supermesh approximation 
in which the cost is proportional to $N_{\rm m} \log(N_{\rm m}) \log(N_{\rm s})$, 
this method may faster than the supermesh approximation for small $N_{\rm m}$ 
because we do not have to calculate ${\cal I}'_{00}$ and ${\cal I}'_{11}$ in 
the point source approximation\footnote{It should be noted that, in 
{\scriptsize START} \citep{START},  the computational time scales with 
$N_{\rm p} \log(N_{\rm s})$, where $N_{\rm p}$ is the number of the SPH 
particles, by utilising the optical depths for SPH particles in the order of 
distance from the radiation source (see \citealt{RSPH} and \citealt{START} for 
more details). This scaling is better than our point source approximation.  
}. 
{Since the surface area of a Str\"{o}mgren sphere is proportional to 
$\dot{N}^{2/3}$ where $\dot{N}$ is photoionization rate 
(see Section~\ref{stroemgren}), 
treating a source group as a point source underestimates the surface area of 
ionized regions.  We will explore this effect in our tests. } 

\begin{table*}
\caption[]{Rates adopted by the code. The lines are, from top to bottom, reference for:
Case B recombination rates (RRB) of $\HII$, $\HeII$, and $\HeIII$ ; 
dielectronic recombination rate (DRR) of $\HeII$; 
collisional ionization rates (CIR) of $\HI$, $\HeI$, and $\HeII$; 
Case B recombination cooling rates (RCRB) of $\HII$, $\HeII$, and $\HeIII$; 
dielectronic recombination cooling rate (DRCR) of $\HeII$; 
collisional ionization cooling rates (CICR) of $\HI$, $\HeI$, and $\HeII$; 
collisional excitation cooling rates (CECR) of $\HI$, $\HeI$, and $\HeII$; 
bremsstrahlung cooling rate (BCR); 
inverse Compton cooling rate (CCR); 
photoionization cross-sections (CS) of $\HI$, $\HeI$, and $\HeII$. }
\label{table:rates}
\begin{tabular}{cccccccccc}
\hline
RRB & DRR & CIR & RCRB & DRCR & CICR & CECR & BCR & CCR & CS \\
\hline 
(4), (5), (4) &  (2) &  (7), (7), (1) & (4), (5), (4) & (3) & (3), (3), (3) & (3), (3), (3) & (4)  & (6)  & (8)\\
\hline 
\end{tabular}
\\
\begin{flushleft}
(1) \citet{abel97}; 
(2) \citet{aldrovandi73}; 
(3) \citet{cen92}; 
(4) \citet{hummer94};
(5) \citet{hummer98};
(6) \citet{ikeuchi86}; 
(7) \citet{janev87}; 
(8) \citet{agnagn}
\end{flushleft}
\end{table*}

\subsection{Non-equilibrium chemistry} \label{sec:chem}

We solve the non-equilibrium chemistry for e,  $\HI$, $\HII$, 
$\HeI$, $\HeII$, and $\HeIII$ implicitly.  
Note that since we employ the on-the-spot approximation, we use `Case B' 
recombination coefficients to calculate recombination rates of $\HII$, 
$\HeII$, and $\HeIII$ throughout this paper. 

Using the optical depth obtained by the methods described in 
Section~\ref{sec:smart} or \ref{sec:start}, the photoionization rates
of $\HI$, $\HeI$, and $\HeII$ in each mesh are given by 
\begin{equation}
\Gamma_i = \sum_\alpha \Gamma_{i, \alpha}, 
\end{equation}
where $\Gamma_{i, \alpha}$ denotes the radiative contribution from a 
radiation source (or a group of radiation sources), $\alpha$, and 
$i = \HI$, $\HeI$, and $\HeII$. 
The contribution from a point-like radiation source, $\alpha$, is represented 
by 
\begin{equation}
\Gamma_{i, \alpha} = \frac{1}{4 \pi h r_\alpha^2} \int_{\nu_i}^\infty 
  \frac{\dm \nu}{\nu} \sigma_i(\nu) L_\alpha(\nu) 
  \exp\left[-\sum_j {\cal N}_{j, \alpha} \sigma_j (\nu)\right], 
\label{eq:photoionization}
\end{equation}
where $\nu_{i}$ is the threshold frequency for the $i$-th species, 
$\sigma_i(\nu)$ is the cross-section of the $i$-th species, and 
$r_\alpha$, $L_\alpha(\nu)$, and ${\cal N}_{i, \alpha}$ are 
respectively the distance between the luminosity centre and the target mesh, the intrinsic 
luminosity of the radiation source (or the group of the sources), 
and the column density of $i$-th species.
The sum in the exponent runs over all three chemical species. 
When all sources have the same spectral shape, i.e. $L_\alpha(\nu) = C_\alpha f(\nu)$, 
we generate a look-up table for each species as a function of column densities: 
\begin{equation}
g_i({\cal N}_k) = \int_{\nu_i}^{\infty}   
\frac{\dm \nu}{\nu} \sigma_i(\nu) f(\nu) 
  \exp\left[-\sum_j {\cal N}_j \sigma_j (\nu)\right]. 
\label{eq:tables}
\end{equation}
In our case, a look-up table for each chemical species becomes three-dimensional 
table. We have confirmed that 20 logarithmic bins for each column density is 
sufficient. By using the look-up tables, the RT calculation is reduced to evaluating 
the column densities. 

Following \citet{anninos97}, we update the densities of each chemical species 
implicitly by using a backward difference formula (BDF). 
The equations to evolve the density of each species can be generally 
written as 
\begin{equation}
\frac{\dm n_i}{\dm t} = C_i (T, n_j) - D_i (T, n_j) n_i, 
\label{eq:chemistry}
\end{equation}
where $n_i = \rho_i / (A_i m_{\rm H})$, $A_i$ is the atomic mass 
number of the $i$-th species, and $m_{\rm H}$ is the proton mass. 
This time $i$ is either e, $\HI$, $\HII$, $\HeI$, $\HeII$ or $\HeIII$. 
The first term of the right-hand side, $C_i$, is the collective source 
term responsible for the creation of 
the $i$-th species. The second term involving $D_i$ represents 
the destruction mechanisms for the $i$-th species and are thus 
proportional to $n_i$. 

Since the timescales for the ionization and recombination differ by 
many orders of magnitude depending on chemical species, 
Eqn.~(\ref{eq:chemistry}) is a stiff set of differential equations. 
In numerically solving a stiff set of equations, implicit schemes are 
required unless an unreasonably small timestep is employed. 
As in \citet{anninos97} we adopt a BDF. 
Discretisation of Eqn.~(\ref{eq:chemistry})  yields 
\begin{equation}
n_i^{t + \Delta t} = \frac{C_i^{t + \Delta t}\Delta t + n_i^t}{1 + D_i^{t + \Delta t} \Delta t}, 
  \label{eq:bdf}
\end{equation}
where all source terms are evaluated at the advanced timestep. 
However, not all source terms can be evaluated at the advanced 
timestep due to the intrinsic nonlinearity of Eqn.~(\ref{eq:chemistry}).  
We hence sequentially update densities of all species in the order of 
increasing ionization states rather than updating them simultaneously; 
We evaluate the source terms contributed by the 
ionization from and recombination to the lower states at the 
advanced timesteps. This method has been found to be very 
efficient and accurate \citep[e.g.][]{anninos97, yoshikawa06}. 

Further improvements in accuracy and stability can be made by 
subcycling the rate solver over a single timestep with which 
the RT is solved. The subcycle timestep, which we call the `chemical timestep',  
is determined so that the maximum fractional change in the electron density  
is limited to 10\% per timestep:  
\begin{equation}
\Delta t_{\rm chem} = 0.1 \left|\frac{n_{\rm e}}{\dot{n}_{\rm e}}\right|. 
\label{eq:chemicaltimespte}
\end{equation}

\subsection{Photo-heating and radiative cooling} \label{sec:heatcool}

Similarly to the photoionization, photo-heating rate for each mesh 
due to the photoionization of the $i$-th species is given by 
\begin{equation}
{\cal H}_i = \sum_\alpha {\cal H}_{i, \alpha}, 
\end{equation}
where ${\cal H}_{i, \alpha}$ indicates the contribution from 
a radiation source (or a group of sources), $\alpha$, and 
$i =  \HI$, $\HeI$, and $\HeII$. 
The total photo-heating rate is defined by ${\cal H} = \sum_i {\cal H}_i n_i$. 
The contribution from a point-like source,
$\alpha$, is written as 
\begin{equation}
{\cal H}_{i, \alpha} = \frac{1}{4 \pi r_\alpha^2} \int_{\nu_i}^\infty 
\frac{\dm \nu}{\nu} \sigma_i(\nu) L_\alpha(\nu) (\nu - \nu_i) 
  \exp\left[- \sum_j \sigma_j(\nu) {\cal N}_{j, \alpha} \right]. 
\label{eq:photoheat}
\end{equation}
As for the photoionization, we generate a look-up table for each species 
when all sources have the identical spectral shape. 

We solve the energy equation for each mesh implicitly as 
\begin{equation}
u^{t + \Delta t} = u^{t} + \frac{{\cal H}^{t + \Delta t} - \Lambda(n_i^{t + \Delta t}, T^{t+\Delta t})}{\rho^t} \Delta t, 
\label{eq:heatcool}
\end{equation}
where $u$ and $T^t = T(n_i^t, u^t)$ are respectively the specific internal energy 
and temperature of the gas  and $\Lambda$ is the cooling function.  
Although this implicit integration is always stable, we need to subcycle the energy 
solver with $\Delta t_{\rm chem}$ because both $C_i$ and $D_i$ in Eqn.~(\ref{eq:bdf})
are functions of the temperature. 
We thus perform the rate solver and the energy solver alternately. 
The chemical timestep $\Delta t_{\rm chem}$ is recalculated before every subcycle.

\subsection{Chemical reaction and cooling rates}

We try to use the chemical reaction and cooling rates 
as up-to-date as possible. The sources of these rates 
are summarised in Table \ref{table:rates}. 
Note that there are notable differences in the recombination 
cooling rates between literatures \citep[see][]{iliev06}.

\subsection{Time stepping} 

Since the optical depth $\tau(\nu)$ at $t + \Delta t$ depends on densities of all 
species at $t + \Delta t$, we have to solve the static RT equation (Eqn.~(\ref{eq:formal2})), 
the chemical reactions (Eqn.~(\ref{eq:bdf})), 
and  the energy equation (Eqn.~(\ref{eq:heatcool})) iteratively. 
We iterate these steps until the relative difference in the electron number density 
becomes sufficiently small: 
$|n_{\rm e}^{(n)} - n_{\rm e}^{(n - 1)}|/n_{\rm e}^{(n)} < \epsilon$, 
where superscripts indicate the number of iterations and we set $\epsilon$ 
to $10^{-4}$ throughout this paper. 
The timestep $\Delta t$, with which we solve the RT equation 
to obtain $\Gamma_i^{t + \Delta t}$ and ${\cal H}_i^{t + \Delta t}$, 
could be much larger than the chemical timestep $\Delta t_{\rm chem}$, 
with which we subcycle the rate and energy solvers. 

We however choose to employ a timestep that is defined by the 
timescale of the chemical reactions: 
\begin{equation}
\Delta t_i = \epsilon_{\rm e} \left|\frac{n_{\rm e}}{\dot{n}_{\rm e}}\right|_i +  
\epsilon_{\rm HI} \left|\frac{n_{\rm HI}}{\dot{n}_{\rm HI}}\right|_i, 
  \label{eq:timestep_i}
\end{equation}
where the second term in the right-hand side prevent the timestep from becoming 
too short when the medium is almost neutral. 
Our choice for $\epsilon_{\rm e}$ 
and $\epsilon_{\rm HI}$ are 0.2 and 0.002, respectively. 
We follow the evolution of the system with the minimum of the timestep defined 
by Eqn.~(\ref{eq:timestep_i}), i.e. 
\begin{equation}
\Delta t = \Delta t_{i, {\rm min}}. 
\label{eq:timestep}
\end{equation}
The timestep $\Delta t$ is therefore only about twice as long as the shortest 
chemical timestep, $\Delta t_{{\rm chem}, {\rm min}}$. 
With this timestep, we find the solutions typically within 3 to 6 iteration steps. 
While we can of course use a longer timestep, 
a longer timestep requires more iterations and the total number of steps 
becomes similar or even larger than the case we employ the timestep defined by 
Eqn.~(\ref{eq:timestep}). With a longer timestep,  the solutions sometimes 
never converge. This timestep is in general much shorter than the timestep 
defined by the Courant timestep criterion and therefore we have to subcycle 
the hydrodynamical timestep with this timestep when we couple the RT with 
the hydrodynamics. 

When optically thick meshes exist, the solutions converge very slowly. 
We thus use smoothed photoionization rates, $\tilde{\Gamma}_i$, and 
heating rates, $\tilde{\cal H}_i$, instead of $\Gamma_i$ and ${\cal H}_i$. 
The smoothed rates for the $i$-th mesh is calculated by using adjacent  
26 meshes, i.e. 27 meshes in total, with a Gaussian kernel of the smoothing 
length $\Delta L$. 
Doing this drastically reduces the number of iterations required to find 
the solutions. 
{
This smoothing may introduce the smearing of the I-fronts especially when 
the spacial resolution is poor. 
While we do not find such an effect in our test simulations as we will 
show later, this can be avoided by applying the smoothing only to optically 
thick meshes as done by \citet{RSPH}. 
}

\section{Test simulations}  \label{sec:tests}

In this section, we describe the tests we perform. 
In order to evaluate the accuracy of our tree-based RT algorithms,  
problems should involve many sources. 
Therefore some of the tests presented are  neither simplest 
nor cleanest. 
All test problems are solved in three dimensions, 
with $128^3$ meshes unless otherwise stated.  

\subsection{Test 1 -- Pure hydrogen isothermal $\HII$ region expansion} \label{stroemgren}

\begin{figure}
\begin{center}
\includegraphics[width=8.0cm]{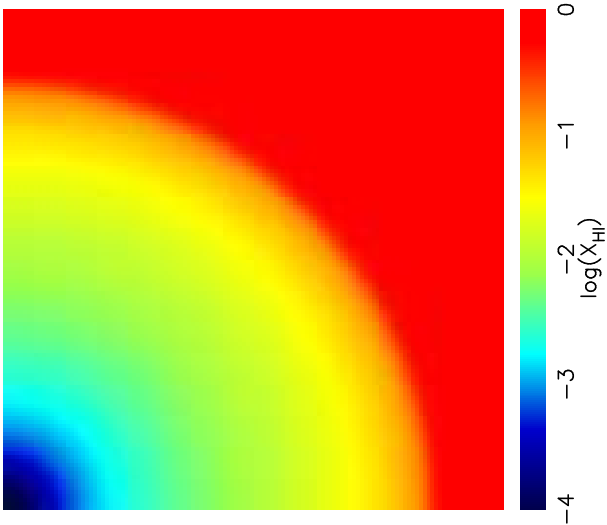}
\end{center}
\caption{Test 1 -- Images of the $\HI$ fraction, cut through 
  at the mid plane of the simulation box at $t = 500~{\rm Myr}$.}
\label{fig:strmap}
\end{figure}

The first test is the classical problem of a $\HII$ region 
expansion in a static, homogeneous, and isothermal gas, which consists of 
only hydrogen, around a single ionizing source.  
This problem has a known analytic solution and is therefore 
the most widely used test. 
Note however that since there is only a single radiation source, 
our RT schemes described in Sections~\ref{sec:smart} and \ref{sec:start} 
have no difference and both methods become the long characteristics method. 
The aim of this test is hence to test our chemical reaction solver and 
time stepping procedure. 

We adopt a monochromatic radiation source that steadily emits 
$\dot{N}_\gamma$ photons per second, whose frequency is the 
Lyman limit frequency ($h\nu_{\rm L} = 13.6$~eV). 
The density of the initially neutral gas is $n_{\rm H}$. 
Assuming the ionization equilibrium,  the Str\"{o}mgren radius 
is given by
\begin{equation}
r_{\rm S} = \left(\frac{3\dot{N}_\gamma}{4 \pi \alpha_{\rm B}(T) n_{\rm H}^2}\right)^{1/3}, 
  \label{eq:stroemgren}
\end{equation}
where $\alpha_{\rm B}$ is the Case B recombination coefficient. 
If we assume that the ionization front (I-front) is infinitely thin, the 
evolution of the I-front radius is analytically given by 
\begin{equation}
r_{\rm I} = r_{\rm S} \left[1 - \exp(-t/t_{\rm rec})\right]^{1/3}, 
  \label{eq:irad}
\end{equation}
where 
\begin{equation}
t_{\rm rec} = \left(n_{\rm H} \alpha_{\rm B}\right)^{-1} 
\label{eq:recombinationtime}
\end{equation}
is the recombination time. 

The analytical solution for the profile of the neutral and ionized fractions 
($X_{\rm HI}(r) = n_{\rm HI}(r)/n_{\rm H}$ and $X_{\rm HII}(r) = n_{\rm HII}(r)/n_{\rm H}$)
can also be calculated \citep[e.g.][]{agnagn} from the equation of the ionization balance 
at radius $r$:
\begin{equation}
\frac{n_{\rm HI}(r)}{4 \pi r^2} \dot{N}_\gamma e^{-\tau(r)}\sigma_{\rm HI}(\nu_{\rm L}) = n_{\rm HII}(r)^2 \alpha_{\rm B}(T),  
\end{equation}
where 
\begin{equation}
\tau(r) = \sigma_{\rm HI}\int_0^r n_{\rm HI}(r') \dm r'. 
  \label{eq:taur}
\end{equation}
The profile of the neutral fraction is thus given by 
\begin{equation}
X_{\rm HI}(r) = \frac {2 + \frac{\dot{N}_\gamma e^{-\tau(r)} \sigma_{\rm HI}}{4 \pi r^2 n_{\rm H} \alpha_{\rm B}} 
  - \sqrt{\left(2 + \frac{\dot{N}_\gamma e^{-\tau(r)}\sigma_{\rm HI}}{4 \pi r^2 n_{\rm H} \alpha_{\rm B}}\right)^2 - 4}}{2}. 
\label{eq:analytic}
\end{equation}
To derive this profile, we ignore the collisional ionization, which is included 
in our simulations. 

The initial physical parameters of this test are the same 
as those of Test 1 in Cosmological Radiative Transfer Comparison 
Project \citep{iliev06}, where the hydrogen number density, $n_{\rm H}$, 
is $10^{-3}~{\rm cm}^{-3}$, the temperature of the isothermal gas 
is $10^4$~K, and ionization rate, $\dot{N}_\gamma$, is $5 \times 10^{48}$ 
photons~s$^{-1}$. Given these parameters and the recombination rate we use, 
$\alpha_{\rm B}(10^4~{\rm K}) = 2.58 \times 10^{-13}~{\rm cm}^3~{\rm s}^{-1}$, 
the recombination time and the Str\"oemgren radius are 
$t_{\rm rec} = 122.6~{\rm Myr}$ and $r_{\rm S} = 5.4~{\rm kpc}$, respectively. 

We employ identical numerical parameters to those in \cite{iliev06}: 
The side length of the simulation box is 6.6~kpc, initial ionization fraction 
is set to $1.2 \times 10^{-3}$, and a radiation source is placed at the 
corner of the box, $(0, 0, 0)$. 
We compare our simulation results to the analytical solution given by 
Eqn.~(\ref{eq:analytic}) which represents the solution at $t = \infty$.

In Fig.~\ref{fig:strmap}, we show the neutral fraction in the $z = 0.5 \Delta L$ 
plane  at $t = 500~{\rm Myr}$, at which point the I-front is close to to the 
maximum radius, i.e. the Str\"{o}mgren radius. 
The $\HII$ region is nicely spherical, though this is not surprising because, 
with a single source, our method is identical to the long characteristics method.  
\begin{figure}
\begin{center}
\includegraphics[width=8.0cm]{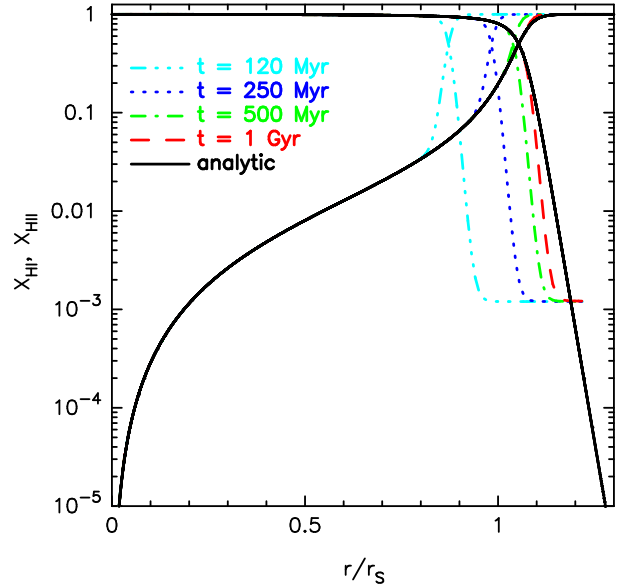}
\end{center}
\caption{Test 1 -- The profiles of ionized and neutral fractions. 
  The radius is in units of the Str\"{o}mgren radius. 
  The dot-dot-dot-dashed, dotted, dot-dashed, and dashed lines represent simulated results at 
  $t = 120$, 250, 500, and 1000 Myr, respectively. 
  The solid line indicates the analytical solution at $t = \infty$ given by Eqn.~(\ref{eq:analytic}).  
  The minimum ionized fraction in the numerical results is set by the collisional ionization which is 
  not included in the analytical solution. 
}
\label{fig:strprofile}
\end{figure}
In Fig.~\ref{fig:strprofile}, we show the profiles of ionized and neutral 
fractions at $t = 120$, 250, 500, and 1000 Myr. 
The results asymptotically approach to the analytical solution at 
$t = \infty$. There is a minimum neutral fraction in the simulation results, 
which is set by the collisional ionization that is not included in the analytical 
solution.  

\subsection{Test 2 -- Pure hydrogen $\HII$ region expansion with thermal evolution}

\begin{figure}
\begin{center}
\includegraphics[width=8.0cm]{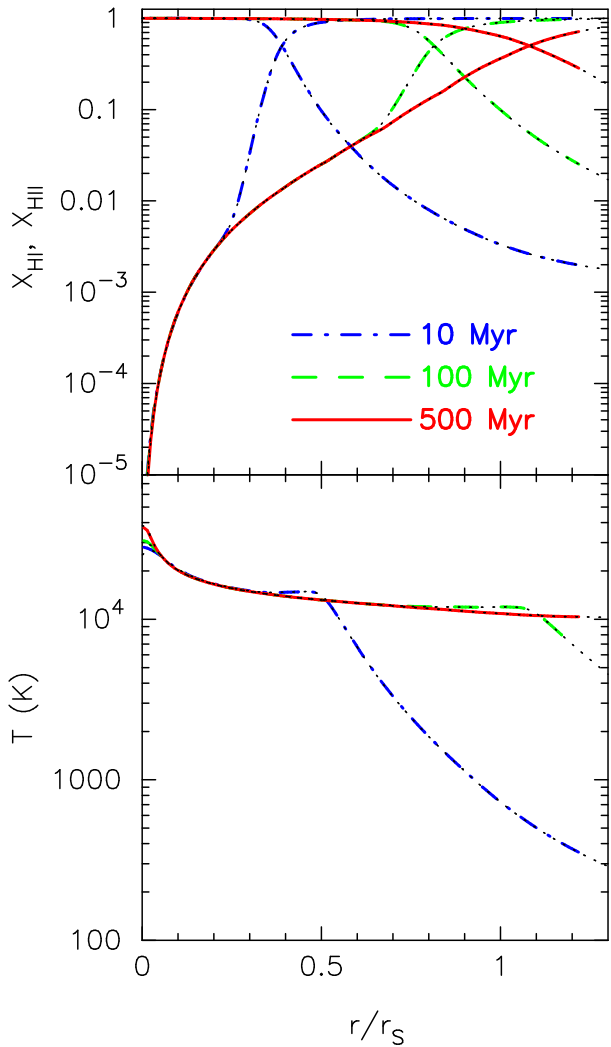}
\end{center}
\caption{Test 2 -- {\it Upper panel}: Spherically averaged ionized and neutral 
  fraction profiles. The dot-dashed, dashed, and solid lines indicate indicate 
  the profile at $t = 10$, 100, and 500~Myr, respectively. 
  The results from a high-resolution spherically symmetric one-dimensional 
  simulation are shown by the dotted lines, which almost perfectly overlap with 
  those by the three-dimensional simulation. 
  The radius is in units of the Str\"{o}mgren radius for the uniform isothermal 
  gas with $n_{\rm H} = 10^{-3}~{\rm cm}^{-1}$ and $T = 10^4$~K. 
  {\it Lower panel}: Spherically averaged temperature profiles. 
  The meaning of the lines is the same as in the upper panel. 
}
\label{fig:strprofilebb}
\end{figure}
Test 2 solves essentially the same problem as Test 1, but the ionizing 
source is assumed to have a $10^5$ K blackbody spectrum and we allow the gas 
temperature to vary owing to heating and cooling processes. 
The initial gas temperature and ionized fraction are 
set to $10^2$~K and $1.2 \times 10^{-3}$, respectively.   

In Fig.~\ref{fig:strprofilebb}, we show the neutral and ionized fraction 
profiles (upper panel) and the temperature profiles (lower panel) at 
$t = 10$, 100, and 500 Myr. 
We also show the results from a high-resolution spherically symmetric 
one-dimensional simulation by the dotted line. For the one-dimensional 
simulation, we use 1024 meshes for a sphere of radius of $1.5 \times r_{\rm S}$ 
and we do not employ the smoothed ionization and heating rates whereas smoothed
rates are employed in the tree-dimensional simulation. 
The results by the three-dimensional simulation are almost indistinguishable 
from those obtained by the one-dimensional one. 
The use of the smoothed rates to accelerate the convergence has thus no evident 
side-effects such as smearing of the I-front. 

{
For this test, our results are most resembling to those obtained by 
{\scriptsize RSPH} for Test~2 in Cosmological Radiative Transfer Comparison Project
\citep{iliev06}\footnote{We  
note that not all codes in Cosmological Radiative Transfer Comparison Project were 
capable of dealing with multifrequency RT. 
}.  
The agreement with {\scriptsize RSPH} is natural because both methods 
are essentially the long characteristics method. Small differences are probably caused by 
different adopted rates. 
}

\begin{figure}
\begin{center}
\includegraphics[width=8.0cm]{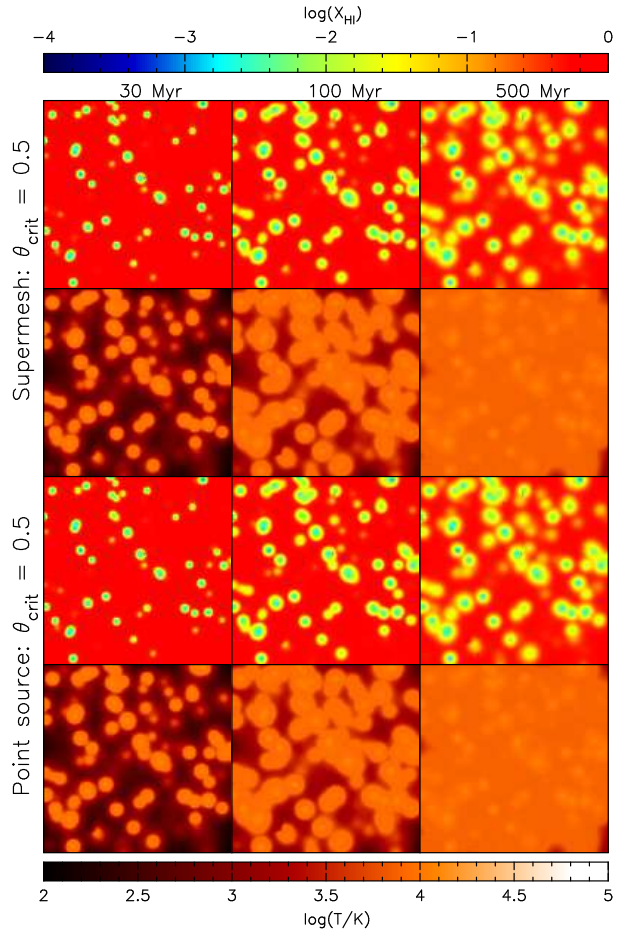}
\end{center}
\caption{Test 3 -- Images of the $\HI$ fraction and temperature, cut through 
  at the mid plane of the simulation box at $t = 30$, 100, and 500 Myr from 
  left to right. The side length of the simulation box is 132 kpc in which we 
  randomly distribute 1000 radiation sources and 1000 optically thick meshes.  
  Upper two rows show results by the supermesh approximation with 
  $\theta_{\rm crit} = 0.5$ and lower two rows by the point source 
  approximation with $\theta_{\rm crit} = 0.5$. }
\label{fig:multiplesources}
\end{figure}

\subsection{Test 3 -- Multiple radiation sources in a clumpy medium} 

In order to test the validity of the RT solver based-on the source 
grouping, we have to solve problems that involve multiple sources. 
Moreover, the error in the supermesh approximation becomes large 
when the inhomogeneity of the medium is large (see Eqn.~(\ref{eq:error})). 
In this test, we therefore solve the RT from multiple sources in 
the clumpy medium. The side length of the simulation box is 132~kpc. 
We randomly select 1000 optically thick meshes whose hydrogen 
number density is $n_{\rm H} = 0.2~{\rm cm}^{-3}$ and optical depth 
at the Lyman limit frequency is $\sim 4 \times 10^3$ for the mesh size. 
The hydrogen number density of other meshes  
is set to $n_{\rm H} = 10^{-3}~{\rm cm}^{-3}$. 
We also randomly distribute 1000 radiation sources in the simulation 
box. Each source has a $10^5$~K blackbody spectrum and steadily 
emits $\dot{N}_\gamma = 5 \times 10^{48}$ ionizing photons per second. 
The initial gas temperature and ionization fraction are set to 
$10^2$~K and $1.2 \times 10^{-3}$, respectively.

\begin{figure}
\begin{center}
\includegraphics[width=8cm]{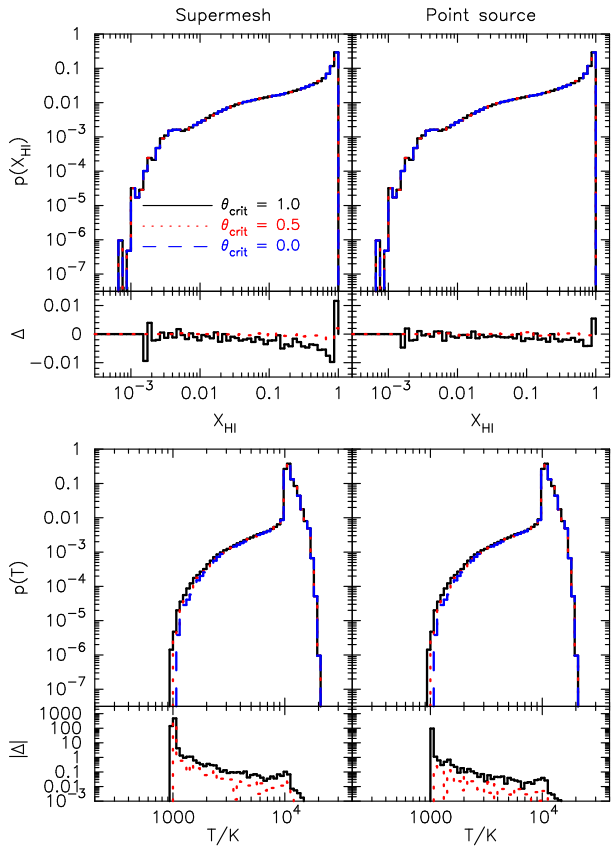}
\end{center}
\caption{Test 3 -- Dependence on the accuracy parameter $\theta_{\rm crit}$.  
  {\it Upper panels}: The volume fractions of the neutral fraction at $t = 500$~Myr. 
  The results by the supermesh approximation are presented in the left panel. 
    The solid (black),  dotted (red), and dashed (blue)  lines indicate the results 
    with $\theta_{\rm crit} = 1.0$, $0.5$, and $0.0$, respectively. 
    The relative difference to the long characteristics method ($\theta_{\rm crit} = 0$), $\Delta$,  
    is also shown. 
  The right panel shows the results obtained by the point source approximation. 
  {\it Lower panels}: The volume fractions of the gas temperature at $t = 500$~Myr. 
  The meaning of the lines are the same as in the upper panels. 
}
\label{fig:multiple_hist}
\end{figure}
\begin{figure}
\begin{center}
\includegraphics[width=8cm]{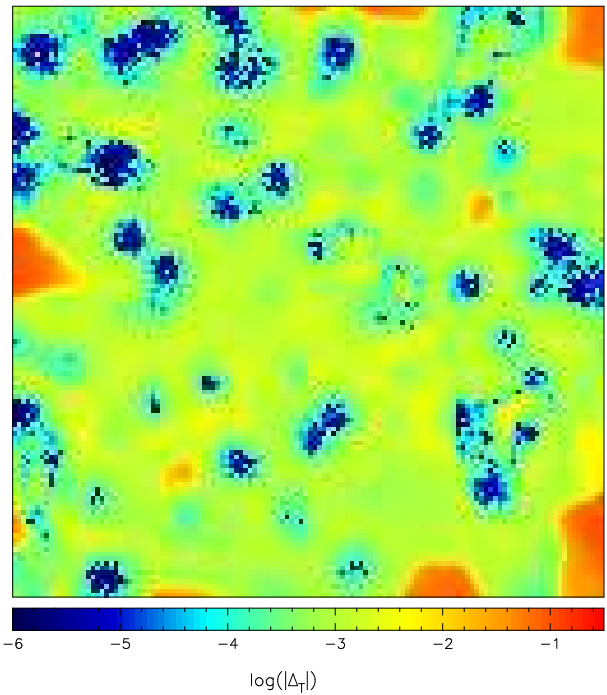}
\end{center}
\caption{Test 3 -- Relative difference in the temperature, cut through 
  at the mid plane of the simulation box at $t = 500~{\rm Myr}$. 
  This figure compares temperature obtained by the supermesh approximation 
  with $\theta_{\rm crit} = 1$ to that by the long characteristics method ($\theta_{\rm crit} = 0$). 
  The relative difference in temperature is defined as 
  $\Delta_T = (\left.T\right|_{\theta_{\rm crit} = 1}^{\rm supermesh} - \left.T\right|^{\rm long}) / \left.T\right|^{\rm long}$. 
}
\label{fig:temp_comp}
\end{figure}
In Fig.~\ref{fig:multiplesources}, we show the neutral fraction and 
temperature maps at the mid plane of the simulation box at $t = 30$, 100, and 
500~Myr.  
We show the results by the supermesh approximation and by the point source 
approximation with $\theta_{\rm crit} = 0.5$. 
The results by two methods are virtually identical to each other including the 
shape of shadows by the optically thick meshes.

In order to investigate the dependence on the accuracy parameter $\theta_{\rm crit}$,  
we compare the simulations with $\theta_{\rm crit} = 1$, 0.5, and 0. 
In Fig.~\ref{fig:multiple_hist}, we show the volume fractions of the neutral 
fraction and the volume fractions of the gas temperature respectively in the 
upper panels and lower panels. 
{We also show difference in the volume fractions relative to those obtained 
  by the long characteristics method ($\theta_{\rm crit} = 0$). 
For example, the relative difference in the volume fraction of the neutral 
fraction by the supermesh approximation with $\theta_{\rm crit} = x$ 
is defined as 
}
\begin{equation}
  \Delta = \frac{\left.p(X_{\HI})\right|_{\theta_{\rm crit} = x}^{\rm supermesh} - 
    \left.p(X_{\HI})\right.|_{\rm long}}{\left.p(X_{\HI})\right|_{\rm long}}.
\end{equation}

{
The volume fractions of the neutral fraction with $\theta_{\rm crit} = 1$ and $0.5$ 
agree quite well with those by the long characteristics method ($\theta_{\rm crit} = 0$). 
The relative differences are typically less than 1~\% even with $\theta_{\rm crit} = 1$. 
For a given value of the accuracy parameter, the point source approximation shows 
slightly better agreement with the long characteristics method. 
On the other hand, agreement in the volume fraction of the gas temperature is 
not as excellent as that for the neutral fraction.  
In particular, both the supermesh and point source approximation predict 
much more low temperature gas around $10^3$~K. 
This is because treating a source group as a point source underestimates the surface 
are of the ionized regions as we stated in Section~\ref{sec:start} and the low temperature gas is primarily heated by 
high energy photons that permeate beyond the surfaces of highly ionized regions. 
Except for this disagreement for the low temperature gas ($ \lesssim 2 \times 10^3$~K), 
typical difference is less than 10~\%. 
}

{
To study how serious the deviation from the long characteristics method at 
low temperature, we compare the temperature map obtained by the supermesh 
approximation ($\theta_{\rm crit} = 1$), which shows the worst agreement with 
the long characteristics method, and that by the long characteristics method 
in Fig.~\ref{fig:temp_comp}. 
We find that the temperature difference is largest for the low temperature 
gas with $T \sim 10^3~{\rm K}$ (see also Fig.~\ref{fig:multiplesources}). 
The difference in temperature is however very small, only 10~\% at most. 
We therefore conclude that the results with $\theta_{\rm crit} = 1$ are almost 
converged to the result obtained by the long characteristics method. 
}

This test proves that both tree-based methods produce equally good results 
even with a large value of the accuracy parameter, $\theta_{\rm crit} = 1$, in 
the situation where a local $\HII$ region is driven primarily by one or a few 
sources. 
This situation is resembling to the early stage of cosmic reionization. 
Only at very late stage of the reionization, the $\HII$ regions overlap each other  
and multiple sources become visible each other; at this stage, 
the reionization has largely completed already. 
We thus expect that our tree-based methods, in particular the supermesh 
approximation, are well suited to this type of problems. 

\subsection{Test 4 -- Clustered radiation sources in a clumpy medium} 

\begin{figure*}
\begin{center}
\includegraphics[width=15cm]{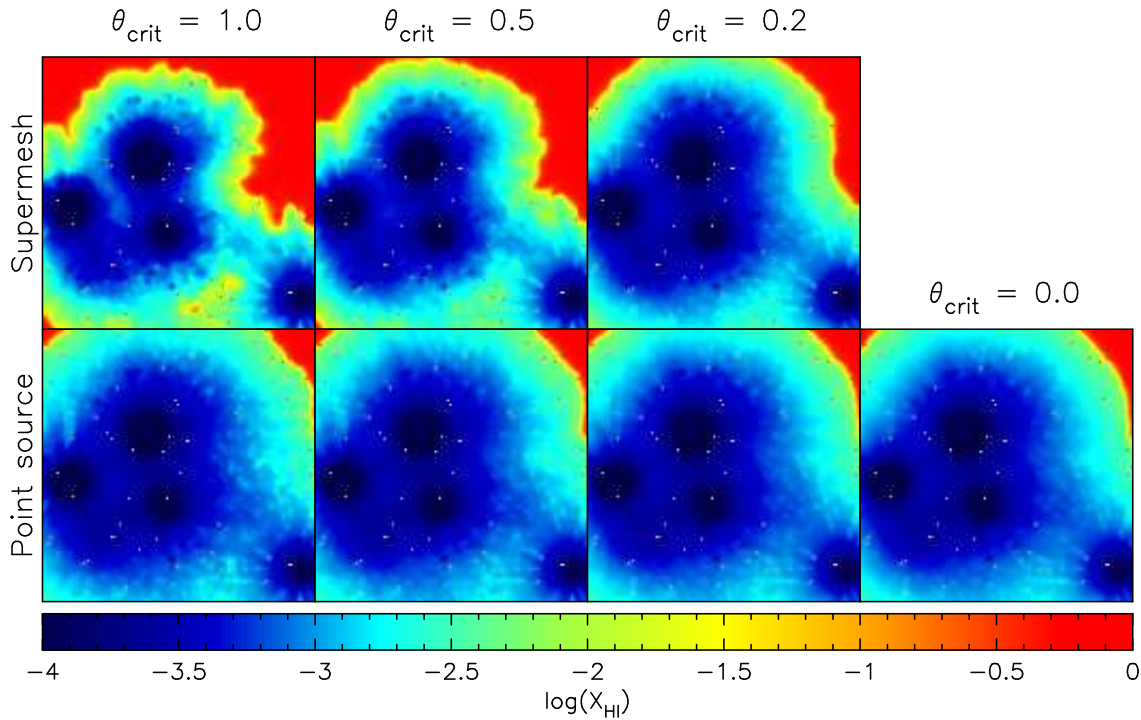}
\end{center}
\caption{Test 4 -- $\HI$ fraction maps, cut through the simulation 
  box at coordinate $z = 63.5 \Delta L = 65.5~{\rm kpc}$ at time 
  $t = 100$~Myr. 
  From left to right, the values of the accuracy parameters are 
  $\theta_{\rm crit} = 1.0$, 0.5, 0.2, and 0.0 respectively.  
  The upper panels show results by the supermesh approximation and 
  the lower panels display those by the point source approximation. 
}
\label{fig:multiple_clustered}
\end{figure*}
Unlike Test 3, here we explore the problem in which groups of sources 
act like bright extended sources and they ionize distant meshes. 
This would be one of the toughest problems for the methods accelerated 
by source grouping. 
The side length of the simulation box is the same as in Test 3, i.e. 
$L_{\rm box} = 132$~kpc. 
In order to construct clustered distribution of radiation sources, 
we put a sphere of radius $r = L_{\rm box}/4$, whose centre is randomly 
placed in the simulation box. 
We uniformly distribute 1000 radiation sources in the sphere. 
We then put a new sphere whose radius is 20\% smaller than the previous 
one and again we distribute 1000 sources in the sphere. 
We continue this procedure until we put 10 spheres, each of which contains 
1000 sources. 
Consequently, there are $10^4$ radiation sources in the simulation box.  
Each source has a $10^5$~K blackbody spectrum and emits 
$\dot{N}_\gamma = 5 \times 10^{48}$ ionizing photons per second. 
We also randomly select $10^4$ optically thick meshes whose hydrogen  
number density is $n_{\rm H} = 0.2~{\rm cm}^{-3}$. 
The hydrogen number density of the remaining meshes is set to 
$n_{\rm H} = 10^{-3}~{\rm cm}^{-3}$. 
The initial gas temperature and ionization fraction are set to 
$10^2$~K and $1.2 \times 10^{-3}$, respectively. 

In Fig.~\ref{fig:multiple_clustered}, we show the neutral fraction maps,  
cut through at the mid plane of the simulation box. 
The size and shape of the ionized regions by the supermesh approximation 
strongly depend on the value of the accuracy 
parameter; The larger the value  is, the smaller the size of the ionized 
regions is. 
This is due to the very nature of the supermesh approximation,
which significantly overestimates the optical depth when a size of 
supermesh is large and the variance of the $\HI$ 
density is large (see Eqn.~(\ref{eq:error}) and (\ref{eq:centrallimit})).  
On the other hand, the results by the point source approximation are 
relatively insensitive to the value of the accuracy parameter. 
The size of the ionized regions is almost same between 
$\theta_{\rm crit} = 1$ and 0 while small difference is seen in the 
shapes.

\begin{figure}
\begin{center}
\includegraphics[width=8cm]{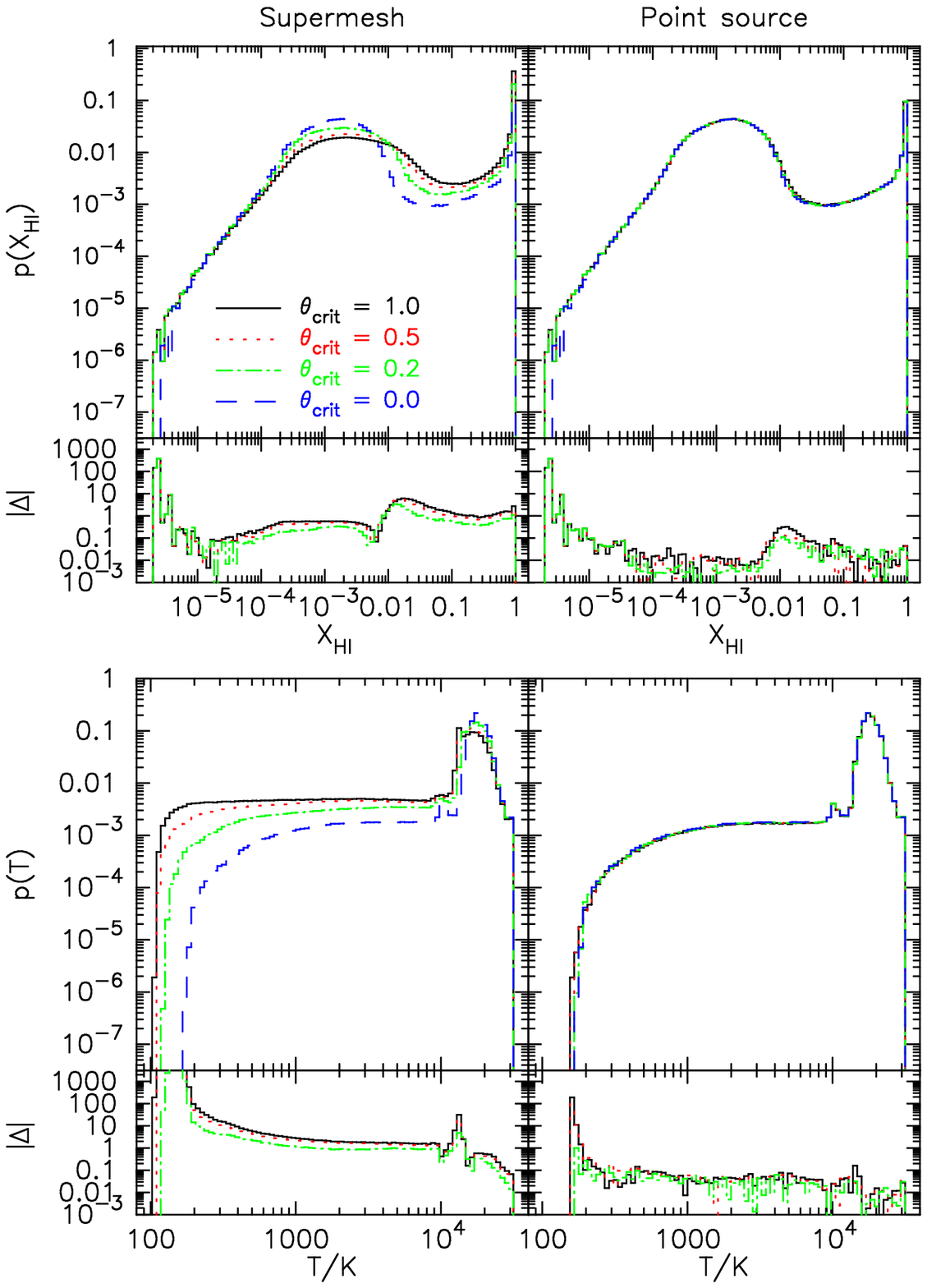}
\end{center}
\caption{Test 4 -- Dependence on the accuracy parameter $\theta_{\rm crit}$. 
  Results at $t = 100$~Myr are presented.     
  The results are displayed in the same manner as Fig.~\ref{fig:multiple_hist}.  
  The volume fractions for the simulation with $\theta_{\rm crit} = 1$, 0.5, 0.2, 
  and 0 are indicated by the solid (black), dotted (red), dot-dashed (green), and dashed (blue) lines, 
  respectively.  
}
\label{fig:clustered_hist}
\end{figure}

In Fig.~\ref{fig:clustered_hist}, we show the volume fractions of the neutral 
fraction and gas temperature at $t = 100$~Myr varying the value of the accuracy 
parameter, $\theta_{\rm crit}$, from 1 to 0. 
We also show the relative difference to the long characteristics method 
($\theta_{\rm crit} = 0$). 
The volume fraction of the neutral fraction confirms the dependence of the 
supermesh approximation on the value of the accuracy parameter, i.e. 
the larger the value of $\theta_{\rm crit}$ is, the smaller the ionized 
fraction is. 
This dependence is more evident in the volume fraction of the gas temperature. 
There is more low temperature gas in the simulation with a larger value of the accuracy parameter. 
Importantly, the results by the supermesh approximation with $\theta_{\rm crit} = 0.2$ 
still significantly deviate from those by the long characteristics methods,  
and therefore we cannot trust the result even with $\theta_{\rm crit} = 0.2$.  

On the other hand, the result by the point source approximation with 
$\theta_{\rm crit} = 1$ shows an excellent agreement with that 
with the long characteristics method, in spite of the fact that this
approximation ignores the spatial extent of source groups. 
This result proves that the point source approximation is very 
efficient and accurate for this type of problems. 

{
The relative difference to the long characteristics method indicates that 
both approximations overestimates the volume fraction of the almost fully-ionized 
gas ($X_{\HI} \simeq 2 \times 10^{-6}$). This ionized fraction corresponds to the  
central regions of each source spheres. 
The volume of these regions are however very small and the neutral fraction is very low 
anyway; this overestimation of the ionization fraction at the central regions of the source 
spheres does not affect the evolution of the whole simulation box. 
In fact, by the point source approximation, the relative difference to the long characteristics method 
in the volume fraction of the neutral fraction is typically 1~\% and $\sim 10$~\% at most except for 
the highly ionized gas with $X_{\HI} \lesssim 10^{-5}$. 
}

{
Even by the point source approximation, 
the relative difference in the volume fraction of the gas temperature to the long characteristics method 
is rather large for the low temperature gas. The gas temperature however 
agrees very well with that by the long characteristics method just as we showed for 
Test~3. Except for the low temperature gas, the typical difference is $\sim 10$~\%. 
Interestingly, decreasing the value of the accuracy parameter in the point source approximation 
from 1 to 0.2 does not improve the agreement with the long characteristics method very much in spite of the 
fact that the simulations with a smaller value of the accuracy parameter is much more computationally 
expensive as we will show in the next subsection. 
Since the point source approximation with $\theta_{\rm crit} = 1$ seems to be sufficiently 
accurate, we expect that this approximation with $\theta_{\rm crit} = 0.5$ would be a safe 
choice for most types of problems. 
}
%

\subsection{Code performance}

We here investigate how the computation time scales with the 
number of meshes and that of the sources. For this purpose, 
we measure the wall-clock time taken for one step of the RT
calculation. 
The computation time for solving chemistry etc. is not included. 
We use 8 cores of 2.13 GHz Xeon E5506 processors for these simulations. 

\begin{figure}
\begin{center}
\includegraphics[width=8.0cm]{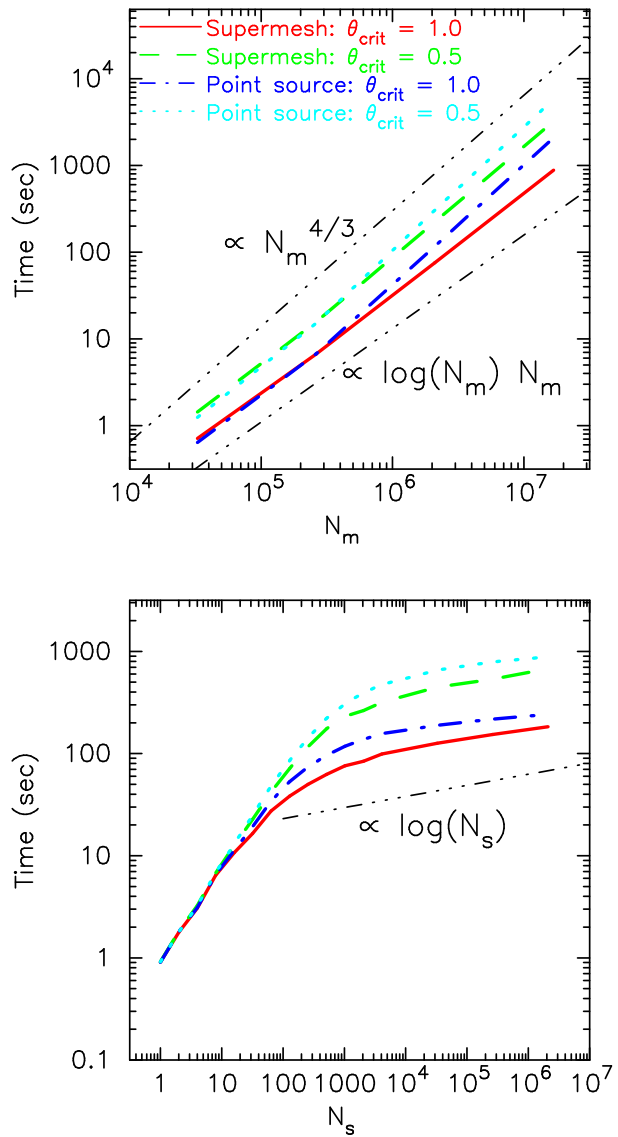}
\end{center}
\caption{Computation time taken for one step of the RT calculation. 
{\it Upper panel}: Computation time as a function of the number of 
meshes, $N_{\rm m}$. The number of radiation sources, $N_{\rm s}$ 
is fixed to 1000. The solid (red) and dashed (green) lines show the 
results by the supermesh approximation with $\theta_{\rm crit} = 
1$ and 0.5, respectively. The dot-dashed (blue) and dotted 
(light blue) lines indicate the point source approximation with 
$\theta_{\rm crit} = 1$ and 0.5, respectively. 
The thin dot-dot-dot-dashed lines show the scaling with 
$N_{\rm m}^{4/3}$ and $N_{\rm m} \log(N_{\rm m})$. 
{\it Lower panel}: Same as the upper panel but the number of the 
source, $N_{\rm s}$, is varied. The number of meshes, $N_{\rm m}$ 
is fixed to $128^3$. The thin dot-dot-dot-dashed line indicates 
the scaling with $\log(N_{\rm s})$. 
}
\label{fig:timing}
\end{figure}

In order to study the scaling with the number of the meshes, we randomly 
place 1000 radiation sources in the simulation box. 
Each source and the simulation box is the same as used in Test 1 except 
that there are 1000 sources and we vary the number of the meshes. 
We show the result in the upper panel of Fig.~\ref{fig:timing}. 
We find that the supermesh approximation is slightly faster than 
the point source approximation for a given set of $N_{\rm m}$ and 
$\theta_{\rm crit}$.  The computation time by 
the point source approximation is a slightly steeper function of 
the number of the meshes than that by the supermesh approximation. 
The computation time by the point source approximation scales with 
$N_{\rm m}^{4/3}$ as expected. 
The scaling of the computation time by the supermesh approximation 
is somewhere between $\propto N_{\rm m} \log(N_{\rm m})$ and 
$\propto N_{\rm m}^{4/3}$. 
Since the RT is solved on the supermeshes whose angular size is similar 
to the angular size of the source group, $\theta_{\rm s}$, which 
can be much smaller than $\theta_{\rm crit}$, the computation time 
becomes steeper function of $N_{\rm m}$ than the expected scaling, 
$\propto N_{\rm m} \log(N_{\rm m})$. 

In the lower panel of Fig.~\ref{fig:timing}, we plot the computation 
time as a function of the number of the sources. The number of the 
meshes is fixed to $128^3$. The computation time scales with 
$\log(N_{\rm s})$ for $N_{\rm s} > 1000$ in all cases. 
This result proves that the tree-based source grouping is quite 
efficient to deal with a large number of radiation sources. 
For a given set of $N_{\rm s}$ and $\theta_{\rm crit}$, 
a simulation by the supermesh approximation is always faster 
than that by the point source approximation. 
It should be however noted that even with the same value of the
accuracy parameter, simulations by the point source approximation 
are sometimes much more accurate than those by the supermesh 
approximation as we showed by Test 4. 

\section{Summary and discussion} 

We have presented a code to solve radiative transfer 
around point sources within a three-dimensional Cartesian 
grid, {\scriptsize ARGOT}, which accelerates the RT calculation
by utilising the oct-tree structure in order to reduce the 
effective number of radiation sources. 
We have explored two methods: one is the supermesh approximation 
and the other is the point source approximation. 
In both methods, sources in a tree node whose angular size is 
smaller than the accuracy parameter $\theta_{\rm crit}$ are 
treated as a single bright source. 
As a result, computation time only scales with $\log(N_{\rm s})$. 
The main difference between these two method is that while the 
former takes the spatial extent of a source group into 
account, the latter ignores the size of the source group and 
treat it as a point source. 
In the supermesh approximation, the RT is solved using supermeshes 
whose angular size is similar to the angular size of the source 
group in question. Doing this results in the further acceleration of 
the RT calculation. 

One might thus see that the supermesh approximation is 
superior to the point source approximation. 
We have however shown that the point source approximation is 
always equally or more accurate than the supermesh approximation 
for a given value of the accuracy parameter. 
This is because RT in a inhomogeneous medium on a supermesh 
inevitably overestimates the optical depth. This approximation can 
be in principle improved by including higher order moments, such 
as variance, although we do not take such an approach. 
This method hence only applicable to the problems in which a 
local $\HII$ region is driven primarily by one or a few sources 
such as Test 3 in this paper. 
When one applies the supermesh method to the simulation of  
cosmic reionization, it could be combined with the `local clumping 
factor' approach proposed by \citet{raicevic11}, although exploring 
such a method is beyond the scope of this paper. 

The point source approximation, which can be regarded as a 
mesh version of {\scriptsize START} \citep{START}, produces 
sufficiently accurate results with $\theta_{\rm crit} = 1$ for 
all test simulations presented in this paper. 
This approximation requires slightly more computational cost 
than the supermesh approximation and it scales with 
$N_{\rm m}^{4/3} \log(N_{\rm s}) $. 
The performance can be improved if we choose the angular resolution 
so that at least one ray from a radiation source (or a group of sources) 
crosses all target meshes instead of solving RT to all target meshes. 
Doing this reduces the total number of rays from $\propto N_{\rm m}$ to 
$\propto N_{\rm m}^{2/3}$. Such an algorithm has been applied for RT from 
point sources \citep{yajima09} and can be extended to our tree-based algorithm. 
The expected scaling is $ N_{\rm m} \log{N_{\rm s}}$, which is even 
faster than the supermesh approximation and the same scaling by 
{\scriptsize START}. 

For parallel implementation, if the entire meshes and sources can fit 
into memory of one computer node, parallelisation via angle decomposition is 
preferable to volume decomposition. We implement the angle decomposition 
by using both {\scriptsize MPI} and {\scriptsize OpenMP}.  
If a simulation size becomes too large to fit into the memory of one computer 
node, we have to employ the volume decomposition. 
The volume decomposition for RT around point sources was introduce by 
\citet{RSPH} and the algorithm can be applied to our methods. 
We leave the volume decomposition to future work. 

The method presented in this paper can be easily combined with 
any grid-based hydrodynamic code, even with codes based on 
AMR \citep{FLASH, RAMSES, ENZO} and 
will be useful for various astrophysical problems in which 
a large number of radiation sources are required such as 
cosmic reionization and galaxy formation. 
We will apply our code for these issues in a forth coming paper.

\section*{Acknowledgements}

We would like to thank Kenji Hasegawa and Hideki Yajima for stimulating 
discussion. We are also grateful to the anonymous referee for helpful comments. 
The simulations were performed with FIRST and T2K Tsukuba at Centre 
for Computational Sciences in University of Tsukub and with the Cray 
XT4 at CfCA of NAOJ. 
This work was supported in part by the {\it FIRST} project based on 
Grants-in-Aid for Specially Promoted Research by MEXT (16002003), 
Grant-in-Aid for Scientific Research (S) by JSPS (20224002). 
TO acknowledges financial support by Grant-in-Aid for Young Scientists 
(start-up: 21840015). 


\begin{thebibliography}{39}
\expandafter\ifx\csname natexlab\endcsname\relax\def\natexlab#1{#1}\fi

\bibitem[{{Abel} {et~al.}(1997){Abel}, {Anninos}, {Zhang} \& {Norman}}]{abel97}
{Abel} T., {Anninos} P., {Zhang} Y., {Norman} M.~L., 1997, \na, 2, 181

\bibitem[{{Abel} {et~al.}(1999){Abel}, {Norman} \& {Madau}}]{abel99}
{Abel} T., {Norman} M.~L., {Madau} P., 1999, \apj, 523, 66

\bibitem[{{Abel} \& {Wandelt}(2002)}]{abel02}
{Abel} T., {Wandelt} B.~D., 2002, \mnras, 330, L53

\bibitem[{{Aldrovandi} \& {Pequignot}(1973)}]{aldrovandi73}
{Aldrovandi} S.~M.~V., {Pequignot} D., 1973, \aap, 25, 137

\bibitem[{{Anninos} {et~al.}(1997){Anninos}, {Zhang}, {Abel} \&
  {Norman}}]{anninos97}
{Anninos} P., {Zhang} Y., {Abel} T., {Norman} M.~L., 1997, \na, 2, 209

\bibitem[{{Aubert} \& {Teyssier}(2008)}]{aubert08}
{Aubert} D., {Teyssier} R., 2008, \mnras, 387, 295

\bibitem[{{Barnes} \& {Hut}(1986)}]{tree}
{Barnes} J., {Hut} P., 1986, \nat, 324, 446

\bibitem[{{Cen}(1992)}]{cen92}
{Cen} R., 1992, \apjs, 78, 341

\bibitem[{{Ciardi} {et~al.}(2001){Ciardi}, {Ferrara}, {Marri} \&
  {Raimondo}}]{ciardi01}
{Ciardi} B., {Ferrara} A., {Marri} S., {Raimondo} G., 2001, \mnras, 324, 381

\bibitem[{{Fryxell} {et~al.}(2000){Fryxell}, {Olson}, {Ricker}, {Timmes},
  {Zingale}, {Lamb}, {MacNeice}, {Rosner}, {Truran} \& {Tufo}}]{FLASH}
{Fryxell} B., {et~al.}, 2000, \apjs, 131, 273

\bibitem[{{Gnedin} \& {Abel}(2001)}]{ga01}
{Gnedin} N.~Y., {Abel} T., 2001, \na, 6, 437

\bibitem[{{Gonz{\'a}lez} {et~al.}(2007){Gonz{\'a}lez}, {Audit} \&
  {Huynh}}]{heracles}
{Gonz{\'a}lez} M., {Audit} E., {Huynh} P., 2007, \aap, 464, 429

\bibitem[{{Hasegawa} \& {Umemura}(2010)}]{START}
{Hasegawa} K., {Umemura} M., 2010, \mnras, 407, 2632

\bibitem[{{Hummer}(1994)}]{hummer94}
{Hummer} D.~G., 1994, \mnras, 268, 109

\bibitem[{{Hummer} \& {Storey}(1998)}]{hummer98}
{Hummer} D.~G., {Storey} P.~J., 1998, \mnras, 297, 1073

\bibitem[{{Ikeuchi} \& {Ostriker}(1986)}]{ikeuchi86}
{Ikeuchi} S., {Ostriker} J.~P., 1986, \apj, 301, 522

\bibitem[{{Iliev} {et~al.}(2006){Iliev}, {Ciardi}, {Alvarez}, {Maselli},
  {Ferrara}, {Gnedin}, {Mellema}, {Nakamoto}, {Norman}, {Razoumov},
  {Rijkhorst}, {Ritzerveld}, {Shapiro}, {Susa}, {Umemura} \&
  {Whalen}}]{iliev06}
{Iliev} I.~T., {et~al.}, 2006, \mnras, 371, 1057

\bibitem[{{Iliev} {et~al.}(2009){Iliev}, {Whalen}, {Mellema}, {Ahn}, {Baek},
  {Gnedin}, {Kravtsov}, {Norman}, {Raicevic}, {Reynolds}, {Sato}, {Shapiro},
  {Semelin}, {Smidt}, {Susa}, {Theuns} \& {Umemura}}]{iliev09}
---, 2009, \mnras, 400, 1283

\bibitem[{{Janev} {et~al.}(1987){Janev}, {Langer} \& {Evans}}]{janev87}
{Janev} R.~K., {Langer} W.~D., {Evans} K., 1987, {Elementary processes in
  Hydrogen-Helium plasmas - Cross sections and reaction rate coefficients},
  {Janev, R.~K., Langer, W.~D., \& Evans, K.}, ed. Springer

\bibitem[{{Krumholz}(2006)}]{krumholz06}
{Krumholz} M.~R., 2006, \apjl, 641, L45

\bibitem[{{Kunasz} \& {Auer}(1988)}]{kunasz88}
{Kunasz} P., {Auer} L.~H., 1988, \jqsrt, 39, 67

\bibitem[{{Mellema} {et~al.}(1998){Mellema}, {Raga}, {Canto}, {Lundqvist},
  {Balick}, {Steffen} \& {Noriega-Crespo}}]{mellema98}
{Mellema} G., {Raga} A.~C., {Canto} J., {Lundqvist} P., {Balick} B., {Steffen}
  W., {Noriega-Crespo} A., 1998, \aap, 331, 335

\bibitem[{{Nakamoto} {et~al.}(2001){Nakamoto}, {Umemura} \&
  {Susa}}]{nakamoto01}
{Nakamoto} T., {Umemura} M., {Susa} H., 2001, \mnras, 321, 593

\bibitem[{{Ohsuga} {et~al.}(2005){Ohsuga}, {Mori}, {Nakamoto} \&
  {Mineshige}}]{ohsuga05}
{Ohsuga} K., {Mori} M., {Nakamoto} T., {Mineshige} S., 2005, \apj, 628, 368

\bibitem[{{O'Shea} {et~al.}(2004){O'Shea}, {Bryan}, {Bordner}, {Norman},
  {Abel}, {Harkness} \& {Kritsuk}}]{ENZO}
{O'Shea} B.~W., {Bryan} G., {Bordner} J., {Norman} M.~L., {Abel} T., {Harkness}
  R., {Kritsuk} A., 2004, ArXiv Astrophysics e-prints:astro-ph/0403044

\bibitem[{{Osterbrock} \& {Ferland}(2006)}]{agnagn}
{Osterbrock} D.~E., {Ferland} G.~J., 2006, {Astrophysics of gaseous nebulae and
  active galactic nuclei}, 2nd edn., {Osterbrock, D.~E.~\& Ferland, G.~J.}, ed.
  University Science Books

\bibitem[{{Pawlik} \& {Schaye}(2008)}]{TRAPHIC}
{Pawlik} A.~H., {Schaye} J., 2008, \mnras, 389, 651

\bibitem[{{Petkova} \& {Springel}(2009)}]{petkova09}
{Petkova} M., {Springel} V., 2009, \mnras, 396, 1383

\bibitem[{{Petkova} \& {Springel}(2011)}]{petkova11}
---, 2011, \mnras, 415, 3731

\bibitem[{Rai{\v c}evi{\'c} \& Theuns(2011)}]{raicevic11}
Rai{\v c}evi{\'c} M., Theuns T., 2011, \mnras, 412, L16

\bibitem[{{Razoumov} \& {Cardall}(2005)}]{razoumov05}
{Razoumov} A.~O., {Cardall} C.~Y., 2005, \mnras, 362, 1413

\bibitem[{{Ricotti} {et~al.}(2002){Ricotti}, {Gnedin} \& {Shull}}]{ricotti02}
{Ricotti} M., {Gnedin} N.~Y., {Shull} J.~M., 2002, \apj, 575, 33

\bibitem[{{Sokasian} {et~al.}(2001){Sokasian}, {Abel} \&
  {Hernquist}}]{sokasian01}
{Sokasian} A., {Abel} T., {Hernquist} L.~E., 2001, \na, 6, 359

\bibitem[{{Stone} {et~al.}(1992){Stone}, {Mihalas} \& {Norman}}]{stone92}
{Stone} J.~M., {Mihalas} D., {Norman} M.~L., 1992, \apjs, 80, 819

\bibitem[{{Susa}(2006)}]{RSPH}
{Susa} H., 2006, \pasj, 58, 445

\bibitem[{{Teyssier}(2002)}]{RAMSES}
{Teyssier} R., 2002, \aap, 385, 337

\bibitem[{{Wise} \& {Abel}(2011)}]{wise11}
{Wise} J.~H., {Abel} T., 2011, \mnras, 414, 3458

\bibitem[{Yajima {et~al.}(2009)Yajima, Umemura, Mori \& Nakamoto}]{yajima09}
Yajima H., Umemura M., Mori M., Nakamoto T., 2009, \mnras, 398, 715

\bibitem[{{Yoshikawa} \& {Sasaki}(2006)}]{yoshikawa06}
{Yoshikawa} K., {Sasaki} S., 2006, \pasj, 58, 641

\end{thebibliography}

\bsp
\label{lastpage}
\end{document}